# The deterioration of materials from air pollution as derived from satellite and ground based observations

John Christodoulakis[1], Costas A. Varotsos[1], Arthur P. Cracknell[2] and George A. Kouremadas[1]

[1] *Climate Research Group, Division of Environmental Physics and Meteorology, Department of Physics, National and Kapodistrian University of Athens, University Campus Bldg. Phys. V, Athens 15784, Greece.*
[2] *Division of Electronic Engineering and Physics, University of Dundee, Dundee, DD1 4HN, UK*

**ABSTRACT**

Dose-Response Functions (DRFs) are widely used in estimating corrosion and/or soiling levels of materials used in constructions and cultural monuments. These functions quantify the effects of air pollution and environmental parameters on different materials through ground based measurements of specific air pollutants and climatic parameters. Here, we propose a new approach where available satellite observations are used instead of ground-based data. Through this approach, the usage of DRFs is expanded in cases/areas where there is no availability of in situ measurements, introducing also a totally new field where satellite data can be shown to be very helpful. In the present work satellite observations made by MODIS (MODerate resolution Imaging Spectroradiometer) on board Terra and Aqua, OMI (Ozone Monitoring Instrument) on board Aura and AIRS (Atmospheric Infrared Sounder) on board Aqua have been used.

## 1. INTRODUCTION

As the world population is growing rapidly, human contribution to environmental impacts has intensified, with manifestations mainly on climate change, depletion of the ozone layer, and severe urban air pollution[1-4]. The corollary of this is global warming, ozone layer thinning, greenhouse effect enhancement and excessive air pollution in many cities[5-14].

This problem becomes complex as the climate system behaves as self-organizing critical system that displays scaling features, i.e. a small fluctuation is followed by another magnified one by power-law (scaling effect) [15-21]. For example, this gives a better insight to the increase in intensity and frequency of extreme atmospheric events [22-26]. The latter is of great importance for accurate information on critical environmental issues and their impacts[27-32]

The role of air pollution, in connection with climatic conditions, in the deterioration of construction materials exposed to the weather has already been recognized in literature[33-35].

In order this effect to be modeled, a tool called the Dose Response Function has been developed. This function quantifies the corrosion or soiling of a given material due to specific air pollutants, like $HNO_3$, $SO_2$, $NO_2$, $PM_{10}$ (Particulate Matter with aerodynamic diameter equal or less than 10 micrometers), and climatic parameters like temperature, humidity and rainfall. For individual materials, their DRFs are developed involving the air pollutants and climatic parameters which affect the particular material. The development of these functions is based on corrosion/soiling experiments made under laboratory conditions but also in the real environment[36-38]. In this preliminary study, we use part of the available DRFs and specifically the ones for the materials carbon steel, zinc, limestone and modern glass, developed in the framework of the project "International Co-operative Programme on Effects on Materials including Historic and Cultural Monuments (ICP Materials)". The common characteristic of these DRFs, but also of those for other materials already presented in the literature, is that they need as input ground based measurements of air pollutants and climatic parameters. This characteristic limits their applicability to regions/cases where the necessary ground based data are available.

Observations made by remote sensing instruments, like environmental satellites, can help in overcoming this restriction as they are available globally. During the last few years a great challenge that the remote sensing community has tried to confront is the development of satellite instruments capable of making precise and accurate observations near the Earth's surface. A remarkable example of this effort is the OMI which combines the characteristics of GOME (Global Ozone Monitoring Experiment), SCIAMACHY (SCanning Imaging Absorption SpectroMeter for Atmospheric CHartographY) and TOMS (Total Ozone Mapping Spectrometer), in order to retrieve observations for several trace gases and improve spatial analysis. TROPOMI is another example of new instrument, prepared for launch in 2017, which is going to improve satellite observations by increasing spatial accuracy and temporal resolution. As this effort goes on, new kind of near-surface observations will become available which are expected to improve and expand the validity of satellite observations. A field where these observations could be used is in the presented DRFs.

This paper focuses on the presentation and evaluation of a new technique which has been developed in order to expand the use of the already existing satellite data by inserting a specific processing of them so that they can be used with the already developed DRFs. In the following section this technique is described in detail, while in section 3 it is evaluated in connection with experimental results for the case of Athens, Greece. Final conclusions and remarks are given in section 4.

## 2. METHOD AND DATA

The DRFs used in this study are presented in the following, along with the material under study [39,40]

**Carbon Steel**

$$ML = 51 + 1.39[SO_2]^{0.6}Rh_{60}e^{f(T)} + 0.593PM_{10} + 1.29Rain[H^+] \qquad (1)$$

$f(T) = 0.15(T-10)$, for $T<10°C$ \qquad (1.1)

$f(T) = -0.054(T-10)$, for $T\geq 10°C$ \qquad (1.2)

**Zinc**

$$ML = 3.5 + 0.471[SO_2]^{0.22}e^{0.018Rh+f(T)} + 1.37[HNO_3] + 0.041Rain[H^+] \qquad (2)$$

$f(T) = 0.062(T-10)$, for $T<10°C$ \qquad (2.1)

$f(T) = -0.021(T-10)$, for $T\geq 10°C$ \qquad (2.2)

**Limestone**

$$R = 4 + 0.0059[SO_2]Rh_{60} + 0.078[HNO_3]Rh_{60} + 0.0258PM_{10} + 0.054Rain[H^+] \qquad (3)$$

**Modern Glass**

$$H = (0.2215[SO_2] + 0.1367[NO_2] + 0.1092PM_{10}) / (1 + (382/t)^{1.86}) \qquad (4)$$

where
$ML$ = mass loss (the difference in specimen's initial mass minus the remaining mass after removing its corroded part), g m$^{-2}$
$R$ = surface recession, μm (absolute values)
$H$ = haze (%)
$t$ = exposure time, days
$Rh$ = relative humidity, % - annual average
$Rh_{60}$ = $Rh - 60$ when $Rh > 60$, 0 otherwise
$T$ = temperature, °C - annual average
$[SO_2]$ = concentration, μg m$^{-3}$ - annual average
$[NO_2]$ = concentration, μg m$^{-3}$ - annual average
$[HNO_3]$ = concentration, μg m$^{-3}$ - annual average
$Rain$ = precipitation amount, mm year$^{-1}$ - annual average
$PM_{10}$ = concentration, μg m$^{-3}$ - annual average
$[H^+]$ = concentration, mg l$^{-1}$ - annual average. The unit for $[H^+]$ is not the normal one (mol l$^{-1}$) used for this denomination and the relation between $pH$ and $[H^+]$ is therefore here $[H^+] = 1007.97 \cdot 10^{-pH} \approx 10^{3-pH}$.

It should be mentioned here that for the case of $[HNO_3]$ the concentration was estimated by Eq. 5.

$$[HNO_3] = 516e^{-3400/(T+273)} ([NO_2][O_3]Rh)^{0.5} \qquad (5)$$

where

$[O_3]$ = concentration, μg m$^{-3}$ - annual average

In order Eqs. 1, 2, 3 and 4 to be used in connection with satellite observations; the latter should be processed to express these parameters' values as close to the ground as possible and in the appropriate concentration units. For achieving these, the following steps are proposed.

For SO$_2$ estimations, we used OMI SO2 Column Amount (Planetary Boundary Layer) OMSO2e v003 data expressed as daily values in DU (DU = Dobson Unit, 1 DU = 2.69×10$^{20}$ molecules/m$^2$). Firstly, monthly means were calculated. In order for SO$_2$ monthly means to be converted into concentration (μg m$^{-3}$), so they can be used in the previous equations, they were divided by the annual mean height of the Planetary Boundary Layer over Athens, according to literature and molecules quantity was expressed in mass term. Afterwards, the annual mean value was calculated.

For NO$_2$ estimations, we used OMI NO2 Tropospheric Column OMNO2d v003 data, expressed as daily values in mole c/cm$^2$. As for the case of SO$_2$, monthly mean values were first calculated. In order NO$_2$ monthly means to be converted into concentration (μg m$^{-3}$) they were divided, as before, by the annual mean height of the Planetary Boundary Layer over Athens. For estimating the part of this concentration which resides in Planetary Boundary Layer it was used the percentage value proposed in literature [41]. Afterwards, the annual mean value was calculated.

For O$_3$ estimations, we used OMI O3 Total Column OMTO3e v003 data expressed as daily values in DU. Using these data, monthly means were calculated. For estimating the part of this quantity which resides in the troposphere, the monthly mean percentages of tropospheric O$_3$ proposed in the literature were used. In order for the tropospheric O$_3$ monthly means to be converted in concentration they were divided by the monthly mean height of the tropopause as obtained by satellite data (AIRSX3STM v006).

The temperature and relative humidity annual mean values for the surface were obtained directly from satellite data (MOD11C3 v005, AIRSX3STM v006, respectively).

In Eqs. 1, 2 and 3 there is a term which expresses the corrosion due to the wet deposition and which makes use of the rain's pH. This parameter is not observed by satellite instruments so there is a concern about it. According to our results obtained using ground based data, only 0.04% of the carbon steel mass loss, 0.01% of the zinc mass loss and 0.02% of the limestone recession is attributed to this term so it can be excluded without introducing significant errors.

Another remark should be made for the case of PM$_{10}$. Satellite instruments do not make observations of this parameter. As there is, until now, no admissible relevant procedure proposed in the literature which uses satellite observations, like for example Aerosol Optical Depth (AOD), in order to express PM$_{10}$ concentration, in the following we make use of ground based measurements. As a case study, we present soiling estimations of modern glass by making use of PM$_{10}$ concentration calculated using satellite Aerosol Optical Depth (AOD) data (MYD08_M3 v6). This material was chosen for this case study as it is very sensitive to PM$_{10}$.

## 3. RESULTS

In this section are presented the estimates obtained by implementing previous mentioned analysis, referred to as "Satellite based", along with experimental deterioration data collected during one year exposures, the periods October to September of the years 2005-2006, 2008-2009, 2011-2012 and 2014-2015 at Athens, Greece, referred to as "Observed", and the estimates of Eqs 1, 2 and 3 using ground based data, referred to as "Ground based". The procedure involved in the experimental determination of the observed deterioration values has already been published[38]. In that work are also included the results from all the experimental exposure periods and ground based estimates shown in Figs. 1-3 except for the period 2014-2015, which are presented for first time. In Fig. 1 we have plotted the mass loss values of carbon steel samples obtained during experimental exposures (Observed) along with the estimates from Eq. 1 using ground based data (Ground based) and satellite data (Satellite based).

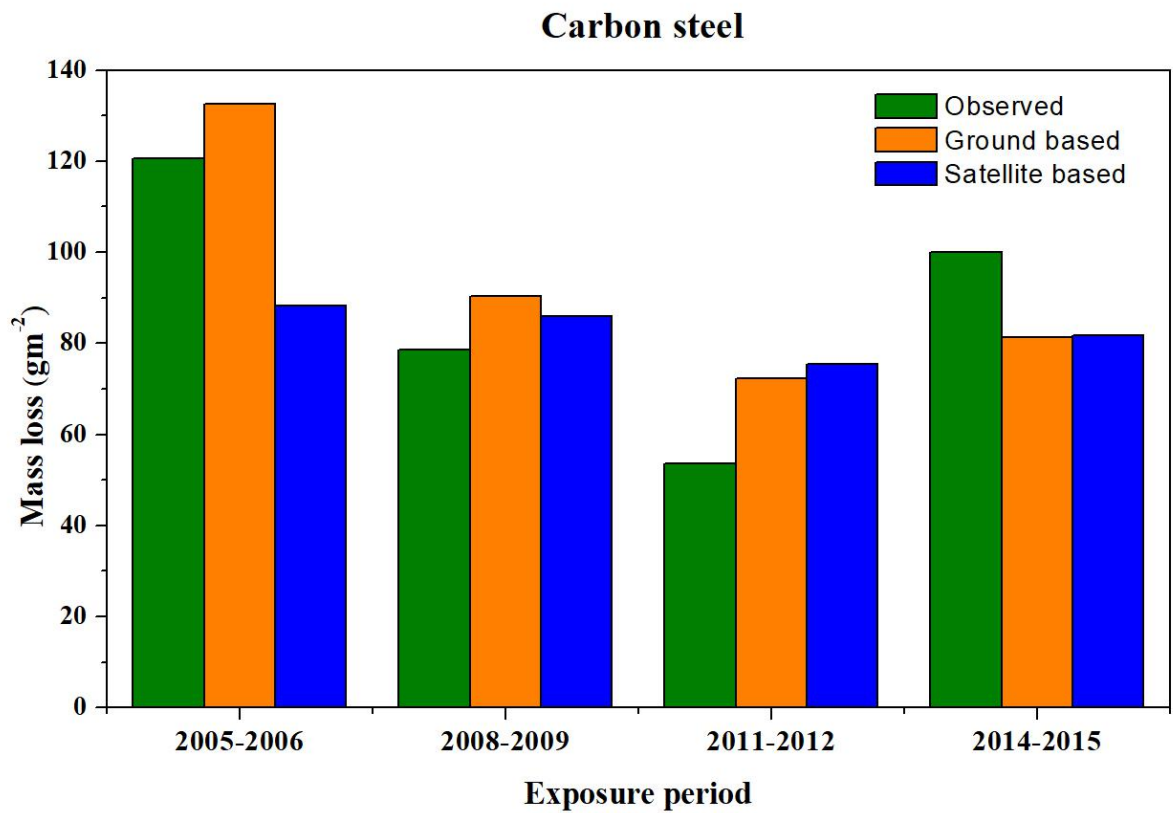

Fig. 1.Mean mass loss of exposed carbon steel samples (observed), carbon steel mass loss estimations using Eq. 1 along with ground based data (Ground based) and carbon steel mass loss estimations using Eq. 1 along with satellite data (Satellite based), for the given periods.

In Fig. 2, are plotted the mass. loss values of zinc samples obtained during experimental exposures (Observed) along with the estimates from Eq. 2 using ground based data (Ground based) and satellite data (Satellite based).

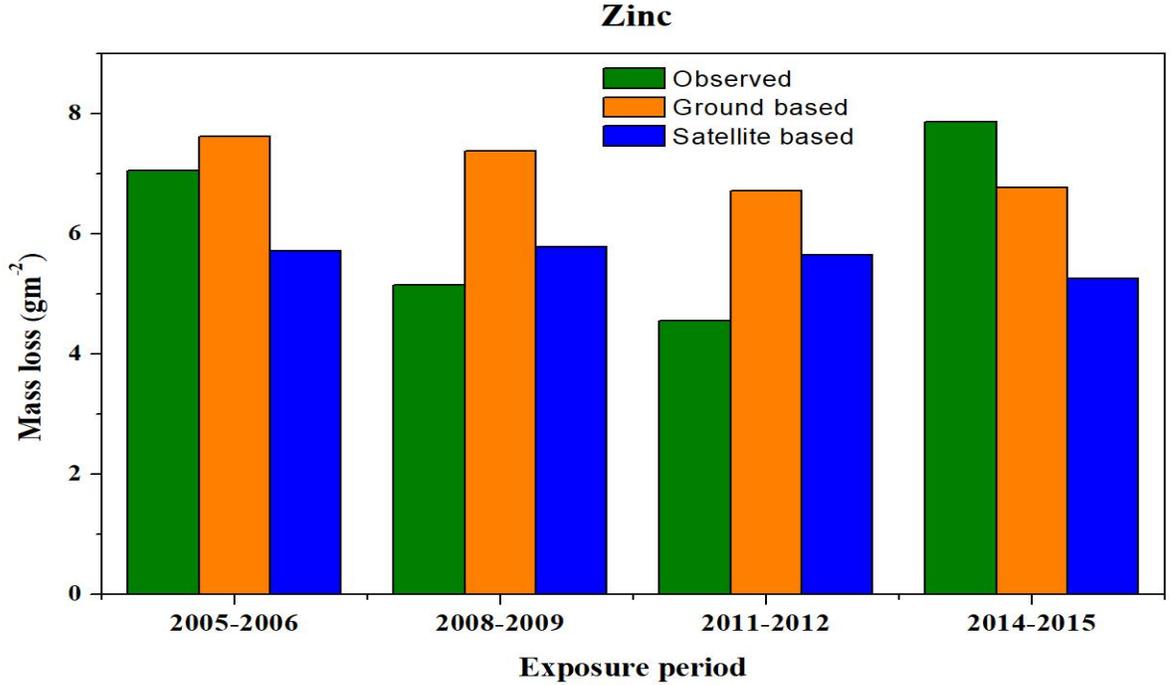

Fig. 2. Mean mass loss of exposed zinc samples (observed), zinc mass loss estimations using Eq. 2 along with ground based data (Ground based) and zinc mass loss estimations using Eq. 2 along with satellite data (Satellite based), for the given periods.

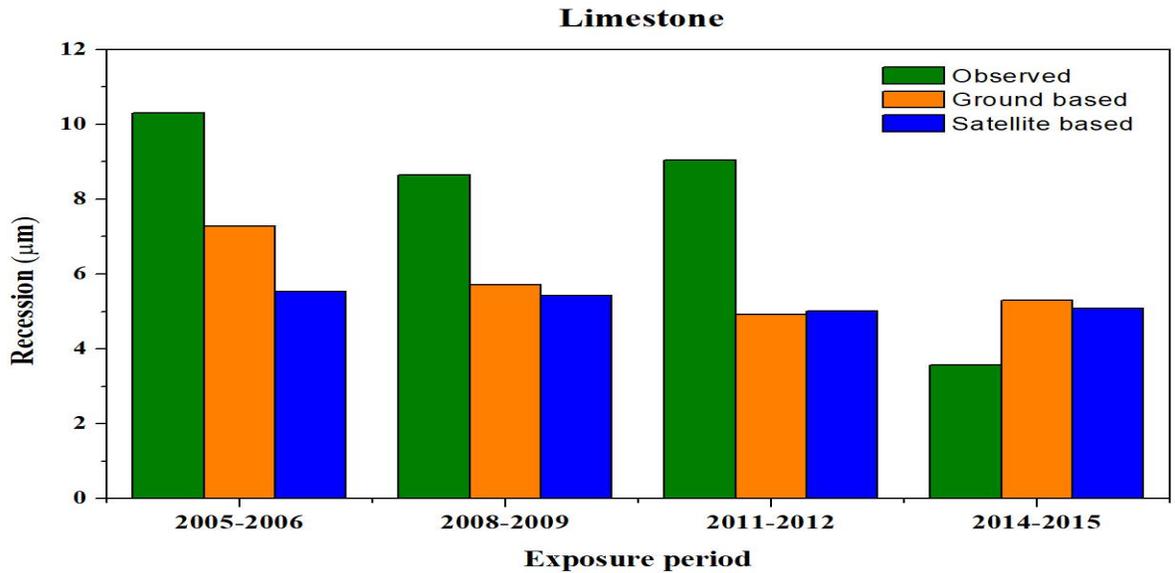

Fig. 3. Mean mass loss of exposed limestone samples (observed), limestone mass loss estimations using Eq. 3 along with ground based data (Ground based) and limestone mass loss estimations using Eq. 3 along with satellite data (Satellite based), for the given periods.

In Fig. 3 we present the recession values of limestone samples obtained during experimental exposures (Observed) along with the estimates from Eq. 3 using ground based data (Ground based) and satellite data (Satellite based).

By inspecting these figures we can examine three questions: i) How well DRFs, using as inputs ground based data, describe the corrosion of the materials? ii) How well DRFs, using as inputs satellite based data, describe the corrosion of the materials? iii) What are the differences between the two methods of estimating corrosion? In order to answer these questions we calculate the followings mean relative errors: i) Mean Relative Error of corrosion estimations obtained using Ground based data to corrosion experimental observations (MRE G/O), ii) Mean Relative Error of corrosion estimations obtained using Satellite based data to corrosion experimental observations (MRE S/O) and iii) Mean Relative Error of corrosion estimations obtained using Satellite based data to corrosion estimations obtained using Ground based data (MRE S/G), respectively. The results are given in Table 1.

Table 1. Mean relative errors among the experimentally observed corrosion levels and the estimations produced by the DRFs using as inputs ground based and satellite based data. MRE G/O = Mean Relative Error of corrosion estimations obtained using Ground based data to corrosion experimental observations, MRE S/O = Mean Relative Error of corrosion estimations obtained using Satellite based data to corrosion experimental observations, MRE S/G = Mean Relative Error of corrosion estimations obtained using Satellite based data to corrosion estimations obtained using Ground based data.

|  | MRE G/O | MRE S/O | MRE S/G |
|---|---|---|---|
| Carbon steel | 10% | 1% | -8% |
| Zinc | 21% | -4% | -21% |
| Limestone | -15% | -21% | -8% |

According to these results, DRFs using ground based data overestimate corrosion levels of carbon steel and zinc by 10 and 21%, respectively, while they underestimate limestone corrosion by 15%. These findings underline the fact that DRFs estimations have limitations to their accuracies. Using the new proposed technique for processing satellite data and implementing them in DRFs the corrosion levels of carbon steel are overestimated by 1%, while for zinc and limestone are underestimated by 4 and 21%, respectively. It is noticed that the corrosion levels of carbon steel and zinc estimated using satellite based data are in better agreement with the experimental observations than the ones using ground based data.

A different way to examine further the validity of the proposed technique is to implement this for mapping the corrosion/soiling estimates. The greater Athens area was chosen as a case study as in this region there are available data for meteorological parameters and air pollutants loads. The corrosion/soiling estimates obtained using Eqs. 1, 2, 3 and 4 for the test period 1 April 2013 to 31 March 2014.

In particular, Figs. 4a, 5a and 6a illustrate the corrosion estimations for carbon steel, zinc and limestone, respectively, using ground based data while Figs. 4b, 5b and 6b illustrate the same parameter using satellite data.

Fig. 7a illustrates the soiling estimations for modern glass based on ground based data while Fig. 7b the soiling estimations using satellite data except for $PM_{10}$ parameter. Fig. 7c illustrates the case study, mentioned earlier in section 2, where $PM_{10}$ concentration has been estimated through AOD observations and the mathematical function, proposed by Pere et al. (2009).

The necessary environmental data were collected from local meteorological stations network of NOA (National Observatory of Athens) while the air pollutants loads were collected from Ministry of Environment network.

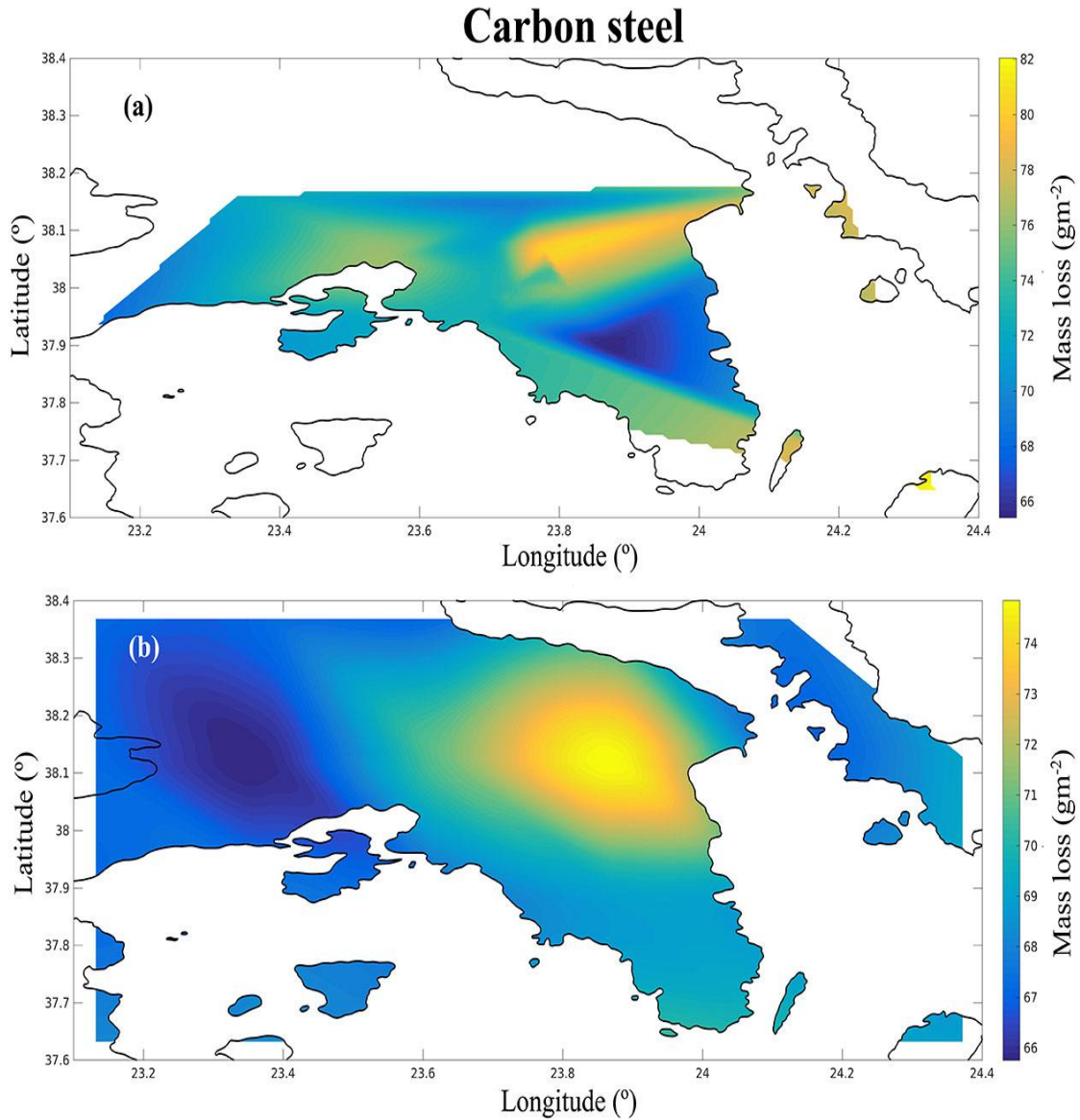

Fig. 4. (a) Mapping of the obtained mass loss estimations of carbon steel by using Eq. 1 with ground based data. (b) Mapping of the obtained mass loss estimations of carbon steel by using Eq. 1 with satellite data.

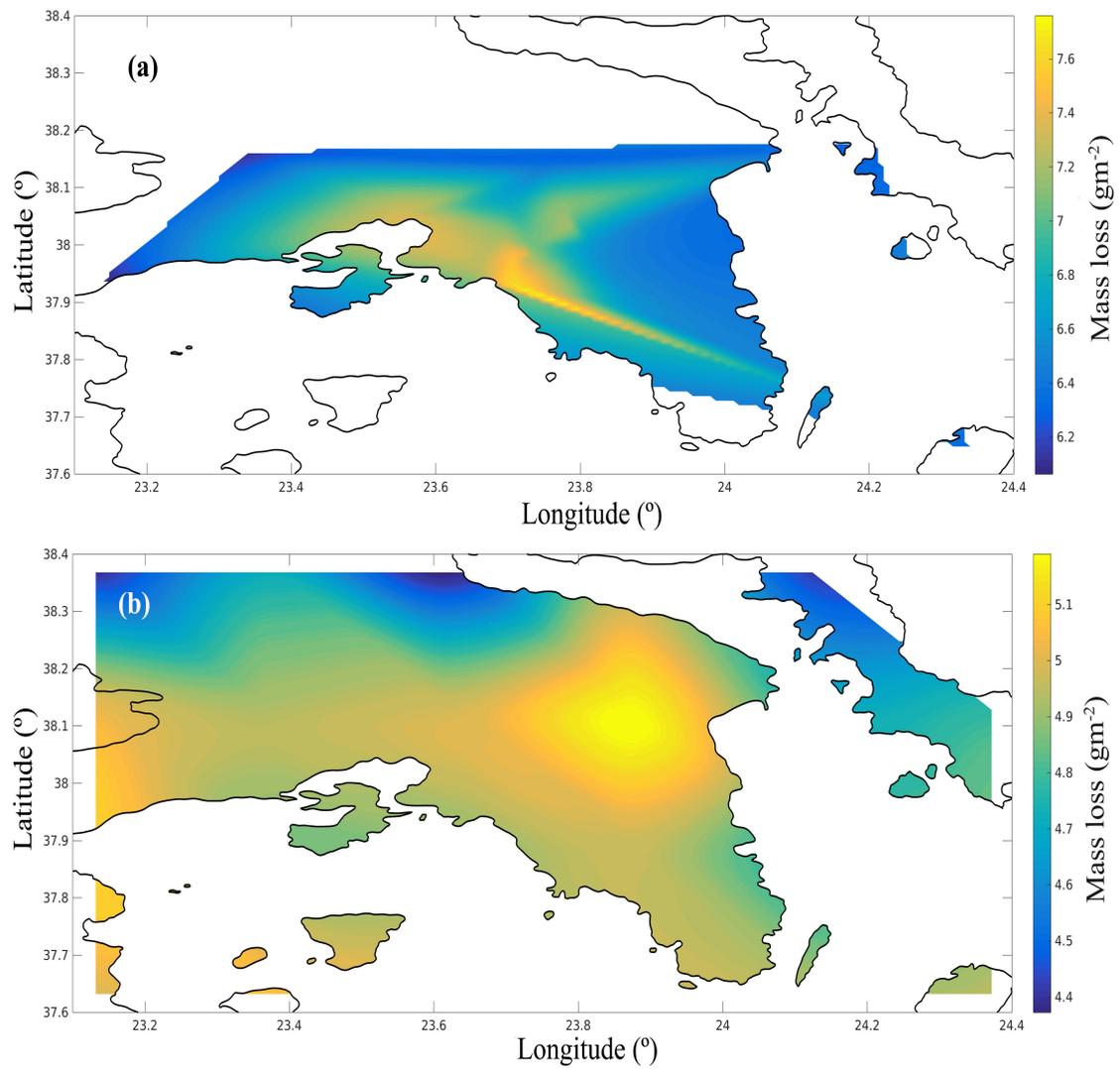

Fig. 5. (a) Mapping of the obtained mass loss estimations of zinc by using Eq. 2 with ground based data. (b) Mapping of the obtained mass loss estimations of zinc by using Eq. 2 with satellite data.

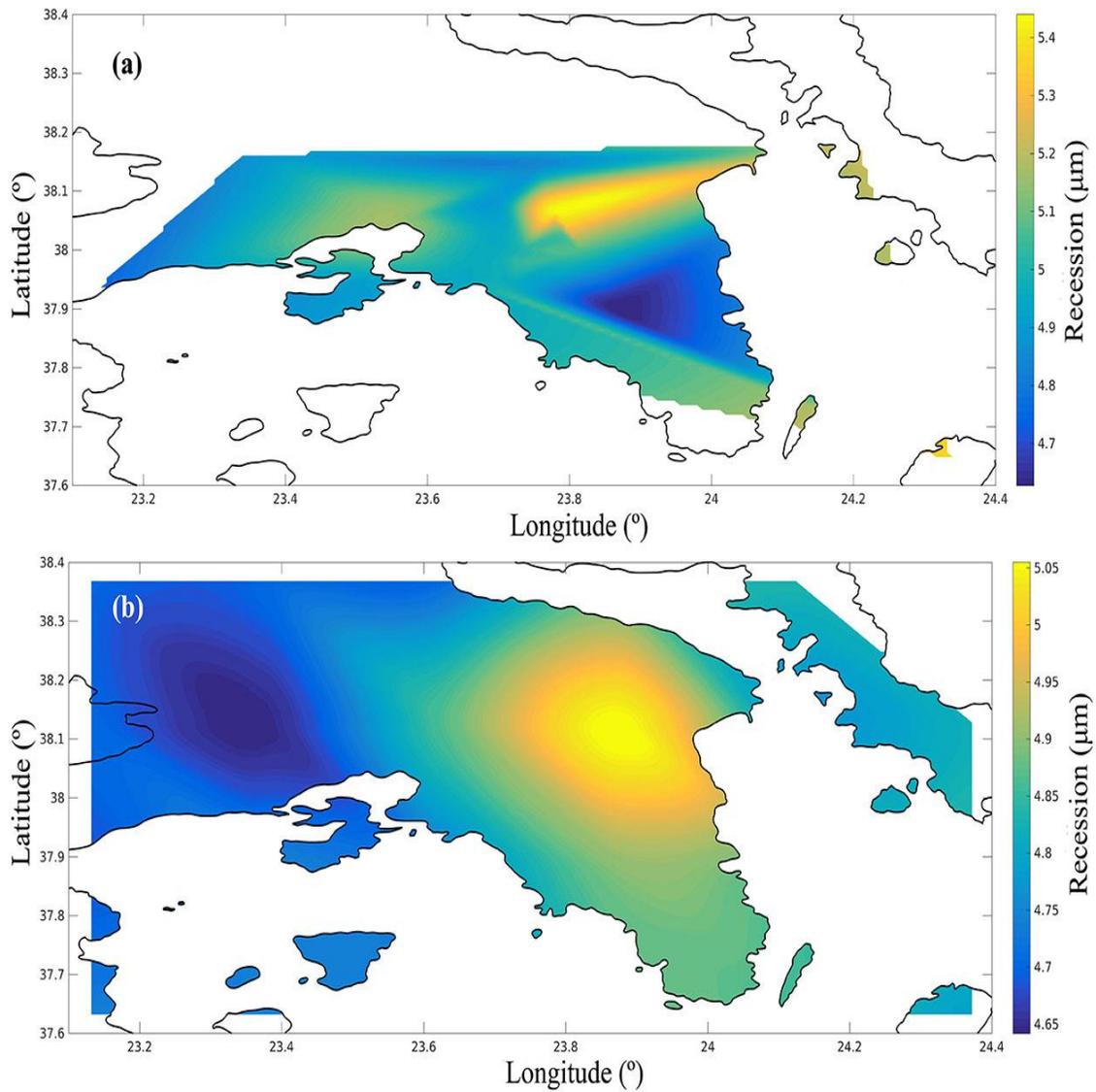

Fig. 6. (a) Mapping of the obtained recession estimations of limestone by using Eq. 3 with ground based data. (b) Mapping of the obtained recession estimations of limestone by using Eq. 3 with satellite data.

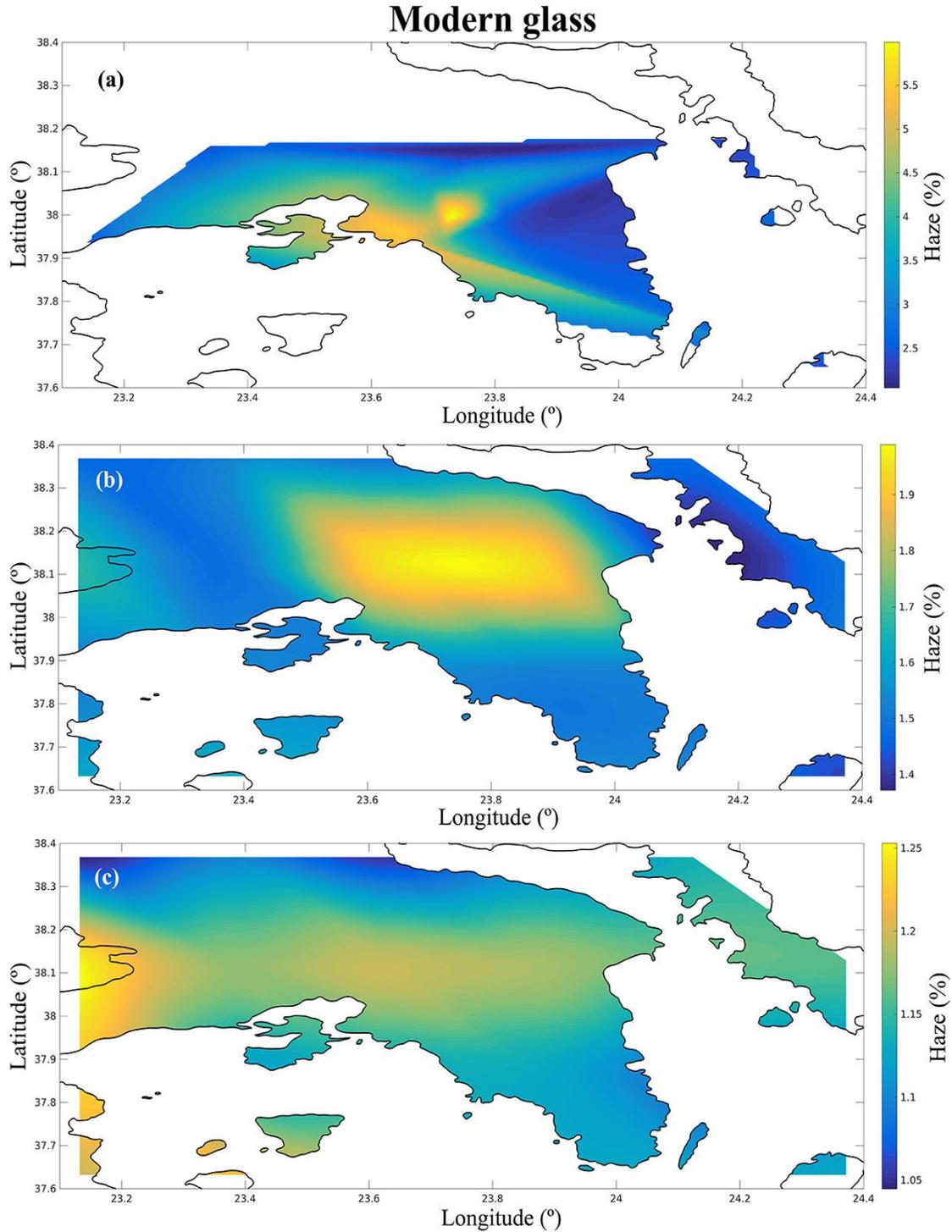

Fig. 7. (a) Mapping of the obtained haze estimations of modern glass by using Eq. 4 with ground based data. (b) Mapping of the obtained haze estimations of modern glass by using Eq. 4 with satellite data, except from $PM_{10}$ concentration. (c) Mapping of the obtained haze estimations of modern glass by using Eq. 4 with satellite data for $PM_{10}$ concentration also.

Qualitative examination of these figures concludes that the presented procedure for processing satellite data in order they can be used in DRFs relying on ground based data, produces results which succeed in locating the regions with high and low values of corrosion/soiling estimations. Implementing spatial data analysis, the following centrographic statistics descriptors were calculated in order to examine previous macroscopic conclusions: Mean Center (MC), Standard Deviation of MC (SDMC), Center of Minimum Distance (CMD), Standard Distance Deviation (SDD) and Standard Deviational Ellipse (SDE). It should be mentioned that for all calculations the corrosion/soiling values of each material were used as weights factors. The obtained results are given in Table 2. According to these, there are small differences among the statistical features of ground based and satellite based distributions. This finding strengthens previous mentioned remark.

Important similarity is observed among corrosion/soiling results obtained using satellite data ("b" figures) while for the case of ground based results two groups could be identified, carbon steel – limestone and zinc – modern glass. An explanation about the spatial distribution differences between these two groups could be the different air pollution and climatic parameters on which their DRFs rely on. Another remark worth mentioning is the scalar between high and low corrosion/soiling values among ground based and satellite based plots. In the first ones steep transitions are observed in almost all cases while nothing similar is observed in the second ones. A possible explanation for this could be relied on the finite satellite instruments' ground resolution as well as the way that retrieval algorithms process their observations.

Table 2. The results of the centrographic statistics descriptors: Mean Center (MC), Standard Deviation of MC (SDMC), Center of Minimum Distance (CMD), Standard Distance Deviation (SDD) and Standard Deviational Ellipse (SDE). In all calculations coordinates have been weighted by the corresponding corrosion/soiling values of the specific material.

| | | Carbon steel | | Zinc | | Limestone | | Modern glass | |
|---|---|---|---|---|---|---|---|---|---|
| | | Ground based | Satellite based | Ground based | Satellite based | Ground based | Satellite based | Ground based | Satellite based |
| **MC** | Lon (°) | 23.819 | 23.816 | 23.810 | 23.811 | 23.818 | 23.817 | 23.772 | 23.807 |
| | Lat (°) | 37.962 | 37.963 | 37.962 | 37.963 | 37.962 | 37.962 | 37.959 | 37.968 |
| **SDMC** | Lon (°) | 0.299 | 0.296 | 0.297 | 0.298 | 0.299 | 0.297 | 0.292 | 0.292 |
| | Lat (°) | 0.127 | 0.125 | 0.124 | 0.125 | 0.126 | 0.125 | 0.119 | 0.125 |
| **CMD** | Lon (°) | 23.827 | 23.828 | 23.814 | 23.820 | 23.827 | 23.829 | 23.772 | 23.812 |
| | Lat (°) | 37.967 | 37.968 | 37.966 | 37.968 | 37.967 | 37.966 | 37.959 | 37.976 |
| **SDD (km)** | | 29.738 | 29.472 | 29.442 | 29.564 | 29.706 | 29.512 | 28.789 | 29.099 |
| **SDE** | Y axis (km) | 36.926 | 36.702 | 36.268 | 36.587 | 36.837 | 36.675 | 34.684 | 36.426 |
| | X axis (km) | 75.575 | 74.846 | 74.963 | 75.193 | 75.517 | 74.985 | 73.671 | 73.806 |
| | θ of Y axis clockwise (°) | 12.976 | 12.790 | 12.779 | 12.686 | 12.831 | 12.737 | 12.589 | 13.029 |

The quantitative analysis of the same figures reveals that for the case of carbon steel, the obtained mass loss estimations produced by using satellite data are in the same range with the ones obtained using ground based data. For the case of zinc, the obtained mass loss estimations produced by using satellite data are about 15% smaller than the ones obtained using ground based data while they also have smaller range than the first ones. For the case of limestone, the obtained recession estimations produced by using satellite data are in the same range with the ones obtained using ground based data. Finally, for the case of modern glass, the obtained mass loss estimations produced by using satellite data, except from $PM_{10}$ concentration, are about 35% smaller than the ones obtained using ground based data. When $PM_{10}$ concentration is estimated by AOD observations and a proposed in the literature mathematical function the soiling estimations are about 45% smaller than the ones obtained using ground based data.

## 4. CONCLUSIONS

The results obtained from the data analysis and previous discussion could be summarized as follows:
- Final DRFs estimations obtained using satellite data are close (about -8%) to the same estimations obtained using ground based data for the cases of carbon steel and limestone, while for the case of zinc this difference is around -20% (Table 1). These results are encouraging with regard the proposed satellite data processing of $SO_2$, $NO_2$ and $O_3$ observations.
- With the proposed satellite data processing approach, DRFs can be used for places where no ground based data are available in order to investigate for areas with enhanced corrosion/soiling levels.
- According to the spatial data analysis results, the obtained corrosion/soiling distributions produced using ground based and satellite based air pollution and climatic parameters observations present similar features.
- More work has to be done for the improvement of the accuracy of the obtained corrosion/soiling estimations using satellite based data. In some cases, like carbon steel and limestone, the discrepancies between ground based and satellite based estimations are small but in other cases, like zinc and modern glass, these discrepancies are important.

**Acknowledgements**
We gratefully acknowledge the Ministry of Health, all the participants in the ICP Materials programme for their support during the implementation of the experimental campaigns. Part of this research was funded by the Special Account for Research Grants of the National and Kapodistrian University of Athens and European Space Agency (ESA Contract No 4000110313/14/I-BG). Satellite data used in this study were produced with the Giovanni online data system, developed and maintained by the NASA GES DISC. We also acknowledge the mission scientists and Principal Investigators who provided the satellite data used in this research effort.

**Note:** <u>This is in the final stage for submission to the Atmospheric Environment journal</u>

**Competing Interests** The authors declare that they have no competing financial interests.

**Correspondence** Correspondence and requests should be addressed C. Varotsos  (email: covar@phys.uoa.gr)